\documentclass{article}
\usepackage{arxiv}
\usepackage[utf8]{inputenc} 
\usepackage[T1]{fontenc}    
\usepackage{hyperref}       
\usepackage{url}            
\usepackage{booktabs}       
\usepackage{amsfonts}       
\usepackage{nicefrac}       
\usepackage{microtype}      
\usepackage{graphicx}       
\usepackage[numbers]{natbib}
\usepackage{doi}
\usepackage{tabularray}
\usepackage{xcolor}

\hypersetup{
    colorlinks=true,
    linkcolor=black,
    urlcolor=black,
}
\graphicspath{{./asset/image/}}

\hypersetup{
    pdftitle={An Open-source Web-based Dataset Labeling Tool for Slice-to-Volume Registration},
    pdfsubject={preprint},
    pdfauthor={Weixun Luo, Alexandre Triay Bagur,},
    pdfkeywords={First keyword, Second keyword, More},
}
\title{SVRDA: A Web-based Dataset Annotation Tool for Slice-to-Volume Registration}

\renewcommand{\undertitle}{}
\renewcommand{\shorttitle}{}
\author{
    \href{https://orcid.org/0009-0008-0646-8575}
    {\includegraphics[scale=0.06]{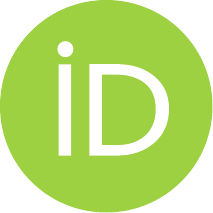}
    \hspace{1mm} Weixun Luo \thanks{This work was conducted when W. Luo was an intern at Perspectum Ltd.}} \\
    Imperial College London \\
    London, UK \\
    \texttt{weixun.luo19@imperial.ac.uk} \\
    \and
    \href{https://orcid.org/0000-0002-5836-737X}
    {\includegraphics[scale=0.06]{asset/image/orcid.pdf}
    \hspace{1mm}Alexandre Triay Bagur} \\
    Perspectum Ltd. \\
    Oxford, UK \\
    \texttt{alexandre.bagur@perspectum.com} \\
    \and
    \href{https://orcid.org/0000-0002-6173-9007}
    {\includegraphics[scale=0.06]{asset/image/orcid.pdf}
    \hspace{1mm}Paul Aljabar} \\
    Perspectum Ltd. \\
    Oxford, UK \\
    \texttt{paul.aljabar@perspectum.com} \\
    \and
    \href{https://orcid.org/0000-0002-0119-050X}
    {\includegraphics[scale=0.06]{asset/image/orcid.pdf}
    \hspace{1mm}George Ralli} \\
    Perspectum Ltd. \\
    Oxford, UK \\
    \texttt{george.ralli@perspectum.com} \\
    \and
    \href{https://orcid.org/0000-0003-0430-0353}
    {\includegraphics[scale=0.06]{asset/image/orcid.pdf}
    \hspace{1mm}Sir Michael Brady} \\
    Perspectum Ltd. \\
    Oxford, UK \\
    \texttt{michael.brady@perspectum.com} \\
}
\date{} 					

\begin{document}
\maketitle
\begin{abstract}
    \item\paragraph{Background and Objective} The lack of benchmark datasets has impeded the development of slice-to-volume registration algorithms. Such datasets are difficult to annotate, primarily due to the dimensional difference within data and the dearth of task-specific software. We aim to develop a user-friendly tool to streamline dataset annotation for slice-to-volume registration.
    \item\paragraph{Methods} The proposed tool, named SVRDA, is an installation-free web application for platform-agnostic collaborative dataset annotation. It enables efficient transformation manipulation via keyboard shortcuts and smooth case transitions with auto-saving. SVRDA supports configuration-based data loading and adheres to the separation of concerns, offering great flexibility and extensibility for future research. Various supplementary features have been implemented to facilitate slice-to-volume registration.
    \item\paragraph{Results} We validated the effectiveness of SVRDA by indirectly evaluating the post-registration segmentation quality on UK Biobank data, observing a dramatic overall improvement (24.02\% in the Dice Similarity Coefficient and 48.93\% in the 95th percentile Hausdorff distance, respectively) supported by highly statistically significant evidence ($p<0.001$).We further showcased the clinical usage of SVRDA by integrating it into test-retest T1 quantification on in-house magnetic resonance images, leading to more consistent results after registration.
    \item\paragraph{Conclusions} SVRDA can facilitate collaborative annotation of benchmark datasets while being potentially applicable to other pipelines incorporating slice-to-volume registration. Full source code and documentation are available at: \href{https://github.com/Roldbach/SVRDA}{https://github.com/Roldbach/SVRDA}.
\end{abstract}
\keywords{Computer Vision \and Medical Image Processing \and Slice-to-Volume Registration \and Dataset Annotation}
\section{Introduction}
\label{section: Introduction}
Image registration is the process by which data from different sources are aligned into a unified coordinate system, constituting one of the cornerstones of computer vision and medical imaging. Particularly in clinical scenarios where multi-dimensional images from diverse modalities or acquisition protocols are frequently cross-referenced for comprehensive decision-making, slice-to-volume registration (SVR), also known as 2D/3D image registration, has been an effective technique that can address the dimensional difference. Various medical applications that rely on such inter-dimensional mapping have emerged, including image fusion for image-guided intervention \cite{Birkfellner2007, Lasowski2008, San2009} and motion artifact correction \cite{Kim1999, Seshamani2013, Gholipour2011}.

However, one of the biggest obstacles hindering the development of SVR algorithms has been the lack of benchmark datasets with gold-standard annotations \cite{Ferrante2017}. These would serve as reliable standards for quantitative evaluation and be indispensable in the supervised learning of deep learning models, which have achieved the state-of-the-art in registration tasks \cite{Haskins2020}. This lack of benchmark datasets for SVR appears to stem primarily from: (1) ground-truth labeling for SVR is inherently more complicated than for other medical image processing tasks such as classification and segmentation. Unlike direct word typing or boundary delineation, SVR labeling entails cross-dimensional resampling of images, which becomes even more challenging when, for instance, 2D slices are acquired obliquely with respect to the reference 3D volumetric scan; and (2) to the best of our knowledge, no task-specific medical image processing software has been developed for labeling benchmark datasets in SVR. This leads to time-consuming annotation with tedious and cumbersome user experience, especially for medical professionals who often have limited computing expertise.

We address these problems by introducing SVRDA, an open-source dataset annotation tool for SVR. SVRDA is entirely web-based and platform-agnostic, enabling users to collaboratively annotate large datasets without fretting over client installation or system compatibility. Leveraging user-friendly keyboard shortcuts, the transformation control mechanism substantially reduces the complexity and latency of manual manipulation. Alongside this, the data-loading mechanism has been tailored to ensure a smooth user experience, enabling seamless transition between cases with auto-saving. Furthermore, numerous intuitive features have been implemented to compensate for the aforementioned labeling difficulties in SVR.

To indirectly validate the effectiveness of SVRDA, we evaluated the quality of post-registration segmentation labels using well-known overlap and surface distance metrics. To showcase the clinical usage of SVRDA, we present a test-retest case study in which the tool was directly integrated into T1 quantification on repeated magnetic resonance images. The results indicate that SVRDA can facilitate accurate and efficient annotation of benchmark datasets for SVR while being potentially applicable to other pipelines embedded with 2D/3D image registration.
\section{Existing Software}
\label{section: Existing Software}
Given the fundamental role of medical imaging in, but not limited to, clinical detection, diagnosis of pathologies, and digital radiology, many related software programs have been developed, each of which can be applied to specific use cases. In this section, we review existing open-source medical image processing software for SVR and discuss their limitations for benchmark dataset annotation.

\paragraph{Low-level Toolkits} Through command-line \cite{Klein2010, Floca2007, Xu2023} or programming \cite{McCormick2014, SimpleITK, Beare2018} interfaces, low-level toolkits allow users to configure various automatic registration algorithms by specifying hyperparameters or freely combining transforms, metrics, optimizers, and interpolators. Based on their highly optimized efficiency and outstanding modularity, low-level toolkits are considered prime algorithmic engines for high-level applications. Unfortunately, such toolkits lack the capability of manual registration. At the same time, their performance heavily relies on the proper selection of the algorithm for the specific context, necessitating either an in-depth algorithmic understanding or a lengthy trial-and-error process from users. These make low-level toolkits difficult to use, notably for clinical specialists lacking extensive computing experience and/or expertise. Additionally, since no graphical widgets are incorporated into such toolkits, extra effort is needed to visualize images, causing inconvenience in ground-truth labeling where outputs must be frequently examined by human annotators.

\paragraph{Graphical Toolkits} To overcome the steep learning curve of scripting and programming required by low-level toolkits, graphical toolkits \cite{FLIRT, Jenkinson2002, AutomatedImageRegistration, Woods1998, Woods1998B} provide simple graphical user interfaces (GUI) for more intuitive interactivity. As data selection and hyperparameter definition are drastically simplified through GUIs, such toolkits are commonly used for algorithm evaluation and tuning. However, there are several residual issues primarily inherited from low-level toolkits, including: (1) no manual registration capability, (2) difficulty in deriving the optimal algorithm for the specific context, and (3) lack of image-display mechanisms. Consequentially, graphical toolkits are fundamentally inadequate for benchmark dataset annotation.

\paragraph{Graphical Applications} Graphical applications \cite{QuickNII, VisuAlign, BioImageSuiteWeb, Fischl2012} offer users the most interactive experience and the broadest range of features including image visualization, acting as complete solutions for their targeted tasks. Beyond computing desired transformation parameters through preset algorithms, users can access more precise and flexible control over them by dragging sliders, clicking buttons, or typing. Nevertheless, in manual fine-tuning where a minor parameter adjustment is frequently followed by several visual inspections, such mechanisms function inefficiently due to the excessive time required to move the mouse across the screen for different operations. In addition, although most graphical applications are able to display 3D volumetric scans from multiple views in 2D, few \cite{3DSlicer, Fedorov2012} can simultaneously render multi-dimensional images in physical space, typically eradicating geometric clues that are extremely beneficial for SVR. Moreover, during sequentially annotating multiple cases in the dataset, users need to explicitly save current outputs and select data for the next case before moving on, resulting in a laborious and error-prone annotation process.
\section{Design Goals}
\label{section: Design Goals}
Recognizing the need for a benchmark dataset annotation tool for SVR, we formulated the following design goals for the tool:

\paragraph{G1: Enable collaboration.} Large-scale annotated image datasets \cite{Deng2009, Lin2014, Kuznetsova2020} have realized tremendous benefits to countless research projects and industrial applications. Nevertheless, building such datasets is most often time-consuming and costly. To address this problem, the tool should empower users to cooperate in dataset labeling, following the success of \cite{VonAhn2004, VonAhn2006, Russell2007}.

\paragraph{G2: Ease annotation.} Acquiring gold-standard annotations in SVR involves sophisticated procedures and requires intensive human efforts in quality control. Moreover, clinical expertise is often necessitated to recognize and match patterns within multi-dimensional representations from disparate medical imaging modalities. To make ground-truth labeling in SVR streamlined and accessible to medical professionals, the tool should be user-friendly and intuitive irrespective of users' technical competence.

\paragraph{G3: Guide annotation.} Throughout annotating ground-truth labels for SVR, the expertise of human annotators profoundly affects the quality of end results. This can lead to high inter-observer variability, which is especially problematic in collaborative labeling. To alleviate this issue, the tool should guide users toward consistently attaining high-quality outputs.

\paragraph{G4: Enable low-latency data visualization.} To provide users with the most information for decision-making in SVR, the tool should include an intuitive image visualization mechanism, which can also assist them in discovering geometric correspondences between data differing in time, space, and modality. Since images must reflect updates in real-time and are repeatedly examined by human annotators throughout the process, this mechanism should guarantee fast responses to user requests as well.

\paragraph{G5: Ease customization.} To facilitate scientific exploitation the tool, it should be flexible and compatible with specific requirements from different research groups and industrial teams with minimal modifications needed. In line with the rapid development of medical image processing, the tool should also be able to incorporate new algorithms smoothly without fundamental redesign.
\section{SVRDA}
\label{section: SVRDA}
SVRDA is designed for annotating benchmark datasets in SVR, with particular application to extracting regions of interest (ROI) in 2D slices using segmentation of a 3D volumetric scan. As a lightweight web-based tool, it can be executed on most contemporary web browsers without laborious installation or setup, serving as a platform-agnostic solution for efficient collaboration (\textbf{G1}). In this section, we describe SVRDA in detail and explain how the design goals in Section \ref{section: Design Goals} have been accomplished in the actual implementation.

\subsection{Prerequisite}
\label{subsection: Prerequisite}
To enable the full functionality of SVRDA, the following data are assumed to be available within a case: (1) one 3D volumetric scan of the patient, referred to as $\mathbf{V}$; (2) one 3D segmentation label marking the ROI(s) in $\mathbf{V}$, referred to as $\mathbf{L}_{3D}$; and (3) a minimum of one 2D slice of the patient, referred to as $\mathbf{S}$.

\subsection{System Overview}
\label{subsection: System Overview}
As shown in Figure \ref{figure: System Overview}, SVRDA is constructed as a four-layer architecture in which each layer is designated for a specific role in the system.
\includeFigure{system_overview}{\useColorTwoPage{An architectural overview of SVRDA.}}{System Overview}

\paragraph{Presentation Layer} This layer consists of the class \textit{AppFactory} and is responsible for handling client-server communication. At compile time, \textit{AppFactory} assembles the GUI and registers callback functions to appropriate widgets. Most of these functions are written in JavaScript and so can be executed on the client, eliminating the lengthy data transmission and page reloading in the request-response communication model.

\paragraph{Business Layer} This layer consists of a collection of independent processing units and is responsible for handling user requests. Each processing unit encapsulates an isolated operation in the registration pipeline as follows:
\begin{itemize}
    \item \textbf{\textit{TransformationProcessingUnit}} computes transformations and stores all the previous outcomes for every $\mathbf{S}$ independently,  referred to as Transformation History.
    \item  \textbf{\textit{ResamplingProcessingUnit}} resamples a 3D image onto the given plane and computes the corresponding 2D cross-section image. By default, trilinear interpolation is used for $\mathbf{V}$ and binary $\mathbf{L}_{3D}$\footnote[1]{The resampled 2D segmentation labels are binarized with a user-adjustable threshold.}, while nearest neighbor interpolation is used for categorical $\mathbf{L}_{3D}$.
    \item \textbf{\textit{MaskingProcessingUnit}} computes all forms of 2D binary image masks.
    \item \textbf{\textit{EvaluationProcessingUnit}} computes the similarity/difference between the given pair of 2D images and determines whether the result is optimal among all the previous outcomes.
    \item \textbf{\textit{PlottingProcessingUnit}} computes all forms of 2D plots.
    \item \textbf{\textit{VisualizationProcessingUnit}} stores style properties (e.g., display windows of images, opacity of 2D segmentation labels, etc.) for all forms of data.
    \item \textbf{\textit{IOProcessingUnit}} saves outputs to the server.
\end{itemize}

\paragraph{Persistence Layer} This layer consists of the class \textit{DataAccessor} and is responsible for managing data within a case. When a callback function is triggered, \textit{DataAccessor} passes the requisite arguments to appropriate processing units through well-defined interfaces and stores the revised data once the communication is finished.

\paragraph{Database Layer} This layer consists of the class \textit{Dataset} and is responsible for managing data within a dataset. Upon the launch of SVRDA, \textit{Dataset} can accept a configuration file and create a corresponding dataset spreadsheet that holds the file locations of all data within the dataset grouped by their case ID. In addition, \textit{Dataset} directly listens to case-shift requests from the GUI and plays a pivotal role in the seamless transition of cases (see Section \ref{subsubsection: Case Transition}).

\subsection{Key Features}
\label{subsection: Key Features}
\subsubsection{Transformation (G2)}
\label{subsubsection: Transformation}
\paragraph{Rigid Transformation} A rigid transformation model with six degrees of freedom (three translation and three rotation parameters) has been deployed, which has shown to be effective in clinical scenarios where there is no apparent image distortion \cite{Gill2008, Jiang2007, Kim2008} and is often utilized to correct major misalignment prior to local deformable analysis \cite{Fuerst2014, Hallack2015, Murgasova2016}. A transformation can be computed in either of two different coordinate systems: (1) a static RAS+ coordinate system, referred to as the Patient Coordinate System; and (2) a dynamic coordinate system based on the real-time pose of $\mathbf{S}$ currently selected by users, referred to as the Slice Coordinate System. This brings substantial flexibility into the control of transformations, easing the handling of $\mathbf{S}$ acquired at oblique angles.

\paragraph{Transformation Control via Keyboard Shortcuts} To simplify manual manipulation, a user-friendly control mechanism has been developed based on keyboard shortcuts. More specifically, pressing \textless W, A, S, D, Q, E\textgreater\space triggers translating $\mathbf{S}$ in the Patient Coordinate System, whereas \textless I, J, K, L, U, O\textgreater\space are for translations in the Slice Coordinate System. Their layout on QWERTY keyboards approximately corresponds to the direction of the eliciting translations, allowing users to effortlessly master the mechanism without strenuous learning. Furthermore, rotations in the Slice Coordinate System can be triggered by pressing Shift + \textless W, A, S, D, Q, E\textgreater, achieving a convenient switch with translations without occupying extra keys. In contrast to the control mechanisms mentioned in Section \ref{section: Existing Software}, ours enables users to manipulate transformations via the keyboard while simultaneously using the mouse to interact with the GUI for other operations, drastically boosting the annotation efficiency throughout the process.

\paragraph{Transformation Control via Buttons} Additional functionalities have been associated with buttons for more productive transformation control as follows:
\begin{itemize}
    \item \textbf{Undo/Redo} enables users to retrieve the previous/next transformation in Transformation History without re-computation, significantly streamlining repeated comparisons between pre and post-registration images.
    \item \textbf{Optimize} enables users to retrieve the transformation that results in the optimal alignment (see Section \ref{subsubsection: Alignment Quantification}) so far without re-computation, usually serving as a better starting point for fine-tuning compared with starting from scratch.
    \item \textbf{Reset} enables users to reset all transformation parameters to zero.
\end{itemize}

\paragraph{Granularity of Transformation Control} The introduction of modes enables users to adjust the granularity of transformation control.
\begin{itemize}
    \item In \textbf{Macro Mode}, all slices in the current case, referred to as $\mathbf{S}_{All}$, are visible and transformations are applied to $\mathbf{S}_{All}$ simultaneously. This is especially useful in multi-slice-to-volume registration where contextual information among $\mathbf{S}_{All}$ should be retained.
    \item In \textbf{Micro Mode}, only the slice currently selected by users, referred to as $\mathbf{S}_{Selected}$, can be seen and transformations are exclusively applied to $\mathbf{S}_{Selected}$. This is frequently utilized in fine-tuning the pose of a particular $\mathbf{S}$.
\end{itemize}

\subsubsection{Case Transition (G2)}
\label{subsubsection: Case Transition}
To facilitate annotation, the tedious and laborious process of setting up new cases has been entirely automated. Once a case-shift request is received, \textit{DataAccessor} first saves the annotated results from the current case to the server. Then, \textit{Dataset} creates a case sub-spreadsheet by extracting all records associated with the new case ID from the dataset spreadsheet. This sub-spreadsheet is passed to \textit{DataAccessor}, where data for the new case can be loaded and pre-processed before being used to update the GUI.

\paragraph{Case Control} \textbf{Previous/Next} buttons enable users to shift to the previous/next case in the dataset, which is designed for sequentially traversing all cases. Also, users can shift to a distant case by selecting its ID in the provided dropdown.

\paragraph{Output Saving} Transformation parameters of $\mathbf{S}_{All}$ can be saved into one Comma-Separated Values (CSV) file and all 2D segmentation labels resampled from $\mathbf{L}_{3D}$, referred to as $\mathbf{L}_{2D}$, can be saved into individual Neuroimaging Informatics Technology Initiative (NIfTI) \cite{Cox2004} files.

\subsubsection{Alignment Quantification (G3)}
\label{subsubsection: Alignment Quantification}
To intuitively guide users throughout the annotation process, the alignment between data is quantitatively monitored in real-time by measuring the similarity/difference between $\mathbf{S}_{Selected}$ and its corresponding 2D cross-section image resampled from $\mathbf{V}$, referred to as $\mathbf{S}_{Resampled}$, using the following evaluation metrics:

\paragraph{Normalized Mutual Information (NMI)} NMI is a similarity metric that measures the ratio of the sum of the marginal entropy and the joint entropy between the given pair of images. Primarily because of its inter-modal robustness, NMI is generally preferred in cases involving multi-modal data.

\paragraph{Sum of Absolute Differences (SAD)} SAD is a difference metric that measures the sum of the absolute difference between pixel values within the given pair of images. It works well in cases where a straightforward correspondence in pixel values between $\mathbf{V}$ and $\mathbf{S}$ is presented.

\subsubsection{Image Visualization (G4)}
\label{subsubsection: Image Visualization}
\paragraph{3D Rendering} Leveraging Dash VTK \cite{DashVTK}, all data within the case are rendered into a unified physical coordinate system, providing users with geometric information to support decision-making in SVR. Various client-side functionalities such as panning, rotating, and zooming have been implemented for swift interaction.

\paragraph{2D Plotting} To promote visual comparisons of data in the feature domain, $\mathbf{S}_{Selected}$ and $\mathbf{S}_{Resampled}$ are plotted in grayscale and overlaid by $\mathbf{L}_{2D}$ with adjustable color and opacity. These four-channel images are compressed into the Portable Network Graphics (PNG) format prior to data transmission for low-latency plotting with lossless quality. Additionally, $\mathbf{S}_{Selected}$ and $\mathbf{S}_{Resampled}$ can be jointly presented using a checkerboard pattern, enabling users to efficiently evaluate the effects of registration visually.

\subsubsection{Masking}
\label{subsubsection: Masking}
Masking has been utilized to restrict operations on certain regions of 2D images. The following masks have been applied to both $\mathbf{S}_{Selected}$ and $\mathbf{S}_{Resampled}$ to enable comparisons:

\paragraph{Positive Mask} This is a static mask only including pixels with positive values in $\mathbf{S}_{Selected}$. During the image reconstruction process, Not-a-Number (NaN) was assigned to pixels that did not receive valid information. Such pixels are replaced by 0 and excluded from resampling, resulting in more consistent plots as shown in Figure \ref{figure: Positive Mask}. Alongside this, the Positive mask can also force more meaningful comparisons between $\mathbf{S}_{Selected}$ and $\mathbf{S}_{Resampled}$ by filtering out their zero-valued backgrounds.
\includeFigure{positive_mask}{\useColorTwoPage{Sample plots with and without applying the Positive mask. (a) The resampled slice without applying the Positive mask. (b) The original slice. (c) The resampled slice after applying the Positive mask. In (a), even though pixels with Not-a-Number values have been replaced by 0 in data pre-processing, they obtain artificial values through interpolation, resulting in confusing features (e.g., the "stretched" tail of the pancreas) highlighted in the plot. Such artifacts have been corrected by applying the Positive mask as shown in (c).}}{Positive Mask}

\paragraph{Overlap Mask} This is a dynamic mask only including pixels in $\mathbf{S}_{Resampled}$ whose physical positions are within $\mathbf{V}$. It effectively restricts evaluation within the overlap between 2D and 3D images, leading to more accurate results by removing outliers from resampling.

\subsection{Flexibility and Extensibility (G5)}
\label{subsection: Flexibility and Extensibility}
\paragraph{Configurable Data Loading} SVRDA accepts configuration files when loading data, where users can set the dataset root directory as well as regular expressions to match the file paths of the desired inputs and synthesize output file locations. This design enables users to work with datasets collected from diverse medical devices or named with distinct standards without modifying source codes.

\paragraph{Plug-in Class Design} To embrace revisions of existing functionalities and introduction of new features by future developers, the classes in SVRDA are designed exploiting a plug-in pattern inspired by \cite{Richards2017} where possible, as shown in Figure \ref{figure: Plug-in Class Design}. These classes consist of a core defining the minimal interface for proper interaction with other classes, and several plug-in modules containing specialized or complex processing operations as private functions to extend/enhance the core. Since those modules have been isolated and are independent of each other, users can safely make module-wise modifications without affecting other parts of the class.
\includeFigure{plug_in_class_design}{\useColorTwoPage{An example class following the plug-in pattern. Only functions in the core (\textit{AppFactory}) can be accessed by other classes for better dependency management.}}{Plug-in Class Design}

\paragraph{Separation of Concerns} To ensure SVRDA is highly extensible, we kept the separation of concerns in mind throughout its development. By encapsulating specific operations into independent classes as discussed in Section \ref{subsection: System Overview}, more complex business logic can be extended by adding new processing units into the Business layer without disrupting existing functionalities. Derived from the architecture of SVRDA where layers are decoupled and communicate with each other through well-defined interfaces, the GUI can be detached and further tailored to users' requirements.
\section{Workflow}
\label{section: Workflow}
In this section, we introduce a typical workflow for dataset annotation to illustrate how users can interact with SVRDA.

\subsection{Uploading the Configuration File}
\label{subsection: Uploading a Configuration File}
To upload a configuration file, users can drag and drop it to the file-uploading widget on the Home page (Figure \ref{figure: Home Page}) or select it through the built-in file browser after clicking the widget. Alternatively, the location of the configuration file can be passed through the provided command-line interface. If the uploaded file cannot be correctly decoded, an error message will pop up.
\includeFigure{home_page}{\useColorTwoPage{The Home page of SVRDA.}}{Home Page}
\includeFigure{main_page}{\useColorTwoPage{The Main page of SVRDA. (A) The Menu section. The Transformation menu (a$_{1}$), 2D Slice menu (a$_{2}$), 2D Resampled Slice menu (a$_{3}$), and 2D Segmentation Label menu (a$_{4}$) can be selected in the dropdown at the top. (B) The Mode section. (C) The Evaluation section. (D) The Support Menu section. The Contour menu (d$_{1}$) and Checkerboard menu (d$_{2}$) can be selected in the dropdown at the top. (E) The Camera section. (F) The 3D Plot section. (G) The Case section. (H) The Main 2D Plot section. (I) The Support 2D Plot section.}}{Main Page}

\subsection{Registering the Slice(s) to the Volume}
\label{subsection: Registering Slices to the Volume}
Once a configuration file is successfully uploaded and decoded, users are automatically directed to the Main page, the main workspace for SVR as shown in Figure \ref{figure: Main Page}.

\paragraph{Controlling transformations} Users can manipulate transformations through keystrokes or clicking the buttons in the Transformation menu (Figure \ref{figure: Main Page}a$_{1}$). To adjust the step sizes used in the keyboard-based transformation control, users can drag the sliders in the same menu.

\paragraph{Controlling the mode} Within the Mode section (Figure \ref{figure: Main Page}B), users can select their preferred mode from the radio button.

\subsection{Evaluating the Alignment}
\label{subsection: Evaluating the Alignment}
Once a transformation has been done, users can both quantitatively and visually evaluate the alignment within the data.

\paragraph{Quantitative Evaluation via Metrics} To measure the similarity/difference between $\mathbf{S}_{Selected}$ and $\mathbf{S}_{Resampled}$, users can select the evaluation metric ("NMI" or "SAD") in the dropdown at the left of the Evaluation section (Figure \ref{figure: Main Page}C). The evaluation result can be hidden by toggling the switch off if preferred.

\paragraph{Visual Evaluation via 3D Rendering} Users can visually evaluate the geometric correspondence between data within the 3D Plot section (Figure \ref{figure: Main Page}F), where both 2D and 3D images are rendered into the same physical coordinate system. By panning, rotating, and zooming the rendered model, users can acquire abundant geometric information facilitating SVR. Also, the pose of the camera can be adjusted by clicking the buttons in the Camera section (Figure \ref{figure: Main Page}E).

\paragraph{Visual Evaluation via Region Fitting} Within the Main 2D Plot section (Figure \ref{figure: Main Page}H), users can check whether $\mathbf{L}_{2D}$ fits in the target region in $\mathbf{S}_{Selected}$, referred to as Region Fitting. To facilitate such examinations, users can zoom a particular area on the plot by dragging a bounding box around it with the mouse for more detailed comparisons. Furthermore, all slices in the case can be individually selected in the dropdown at the top left, while the type ("Resampled Label" or "Resultant Mask") and format ("Contour", "Mask" or "None") of $\mathbf{L}_{2D}$ can be selected in the dropdowns at the top middle and top right, respectively. If $\mathbf{L}_{2D}$ is plotted as a contour, the line width can be adjusted by dragging the slider in the Contour menu (Figure \ref{figure: Main Page}d$_1$).

\paragraph{Visual Evaluation via Boundary Matching} Additionally, users can check whether the boundaries and features from $\mathbf{S}_{Selected}$ and $\mathbf{S}_{Resampled}$ match, referred to as Boundary Matching, in the Support 2D Plot section (Figure \ref{figure: Main Page}I). Most interactions are the same as those in the Main 2D Plot section, except that users can select the type of image ("Resampled Slice" or "Checkerboard") in the dropdown at the top left. If $\mathbf{S}_{Selected}$ and $\mathbf{S}_{Resampled}$ are jointly plotted using a checkerboard pattern, users can adjust the checker width by dragging the slider in the Checkerboard menu (Figure \ref{figure: Main Page}d$_{2}$).

\subsection{Adjusting the Appearance}
\label{subsection: Adjusting the Appearance}
To adjust the appearance of $\mathbf{V}$, users can use the toolbox at the top left of the Plot 3D section (Figure \ref{figure: Main Page}E). The style properties of $\mathbf{S}_{All}$, $\mathbf{S}_{Resampled}$, and $\mathbf{L}_{2D}$ can be adjusted by manipulating the widgets in the 2D Slice menu (Figure \ref{figure: Main Page}a$_{2}$), 2D Resampled Slice menu (Figure \ref{figure: Main Page}a$_{3}$) and 2D Segmentation Label menu (Figure \ref{figure: Main Page}a$_{4}$), respectively.

\subsection{Shifting Cases and Saving}
\label{subsection: Shifting Cases}
After finishing the annotation of the current case, users can shift to the previous/next case by clicking the corresponding buttons in the Case section (Figure \ref{figure: Main Page}G). To shift to a specific case, users can select its ID in the dropdown at the right of the same section. In addition to the auto-saving of annotated results at every case shift, users can manually save them to the server by clicking the Save button at any time.

\subsection{Navigating Pages}
\label{subsection: Controlling the Page}
Users can navigate to pages ("Home" and "Main") by clicking the navigation bar at the top left of SVRDA. An instructional message will pop up once users click "Help" (Figure \ref{figure: Help Modal}), serving as a quick guideline/reminder at runtime.
\includeFigure{help_modal}{\useColorTwoPage{The pop-up message after clicking the "Help" navigation bar. More contents can be accessed by scrolling down the window.}}{Help Modal}
\section{Case Studies}
\label{section: Case Studies}
In this section, we present two case studies to demonstrate the efficacy and clinical usage of SVRDA. All registration in the studies was carried out manually following a protocol based on our prior work \cite{Bagur2021}, as outlined in Appendix \ref{subsection: Registration Protocol}.

\subsection{Indirect Validation via Segmentation}
\label{subsection: Indirect Validation via Segmentation}
To indirectly validate whether SVRDA can be used to annotate high-quality labels, we evaluated the agreement between the post-registration 2D segmentation labels and corresponding ground-truth delineations, which were directly conducted on the slices by ATB, who has over 5 years of experience in medical imaging annotation. The segmentation labels resampled using raw header information were also compared with the manual delineations to construct the baseline.

\paragraph{Dataset} We selected 39 magnetic resonance imaging (MRI) subsets of nominally healthy volunteers from the UK Biobank \cite{Littlejohns2020} where misalignment was evident. By generating the 3D segmentation labels of both the liver and pancreas for each subset, a dataset of 78 cases was obtained, each of which contains:
\begin{itemize}
    \item one neck-to-knee multiple dual-echo Dixon Volumetric Interpolated Breath-hold Examination (VIBE) scan (named “Dixon technique for internal fat - DICOM”, Data-Field ID 20201 on the UK Biobank Showcase), which was acquired over the course of 6 minutes to provide water/fat separation, as $\mathbf{V}$.
    \item one $\mathbf{L}_{3D}$ computed on $\mathbf{V}$ using the implementation of 3D U-Net described in \cite{Owler2021}.
    \item one T1 map derived from the Shortened Modified Look-Locker Inversion recovery (ShMoLLI) scan (named “Pancreas Images - ShMoLLI”, Data-Field 20259 on the UK Biobank Showcase), which was acquired using a proprietary algorithm \cite{Mojtahed2019} from Perspectum Ltd, as $\mathbf{S}$. The T1 map was carefully reviewed to ensure that the liver and pancreas were recognizable.
\end{itemize}
More detailed parameters for data acquisition can be found in Appendix \ref{subsection: Data Acquisition}.

\paragraph{Evaluation Metrics} We applied the Dice Similarity Coefficient (DSC) to effectively assess the overlap between candidate and ground-truth segmentation labels. To gain more insights into locations and deal with labels of irregular structure (i.e., the pancreas), we also used the Hausdorff Distance (HD) for its superior sensitivity to positions, while only taking the 95th percentile of the distance to alleviate the biases from outliers.

\paragraph{Statistical Tests} We utilized the Shapiro-Wilk test to check the distribution normality of the difference between the segmentation quality with and without registration, followed by the two-sided paired Student's \textit{t}-test to check whether the difference was statistically significant. If the normality criterion cannot be satisfied, the Wilcoxon signed-rank test was used as an alternative. We set the significant level to 0.05 in all statistical tests.

\paragraph{No Registration \textit{vs.} After Registration} Figure \ref{figure: Experiment 1 Bar Plot} and Table \ref{table: Quantitative Comparisons} show the results, demonstrating that SVRDA can be used to accurately correct the misalignment within the given data. More specifically, the overall quality of post-registration segmentation labels is substantially improved (by 24.02\% in DSC and 48.93\% in 95\% HD, respectively) compared with the baseline, supported by highly statistically significant evidence ($p<0.001$) in all comparisons. It can also be observed that the raw liver labels show a marginally higher quality than the raw pancreas labels. This results primarily from the rounded shape and larger mass of the liver, which make it more robust than the pancreas to the misalignment induced by the diaphragm movement during breathing. However, even though all cases were manually annotated and delineated under the same conditions, a discrepancy can be found in the post-registration segmentation quality between the liver and pancreas. We contend that such differences are due mainly to the irregular shape and size of the pancreas, which render it vulnerable to motion artifacts and lead to harder manual registration and delineation simultaneously, as shown in Figure \ref{figure: Experiment 1 Example}. This also suggests that target-specific registration protocols might be needed for more consistent results when working with different organs.
\includeFigure{experiment_1_bar_plot}{\useColorTwoPage{Graphical comparisons of the segmentation quality with and without registration evaluated by DSC (left) and 95\% HD (right). Mean $\pm$ 95\% Confidence Interval is presented.}}{Experiment 1 Bar Plot}
\begin{table}[!ht]
    \centering
    \begin{tblr}{
        column{even} = {c},
        column{3} = {c},
        column{5} = {c},
        cell{1}{2} = {c=2}{},
        cell{1}{4} = {c=2}{},
        hline{1,3,6} = {-}{},
    }
                  & DSC (a.u.)       &                          & 95\% HD (mm)                &                                                          \\
                  & No Registration  & After Registration       & No Registration             & After Registration \\
        Liver (N=39)     & 0.7716 ± 0.0445  & \textbf{0.8570 ± 0.0349\textsuperscript{*}}            & 16.26 ± 7.48  & \textbf{10.93 ± 8.23\textsuperscript{*}} \\
        Pancreas (N=39)  & 0.4582 ± 0.0416  & \textbf{0.6682 ± 0.0350\textsuperscript{\textdagger}}  & 31.44 ± 6.77  & \textbf{13.43 + 5.50\textsuperscript{*}} \\
        Combined (N=78)  & 0.6149 ± 0.0463  & \textbf{0.7626 ± 0.0324\textsuperscript{*}}            & 23.85 ± 5.29  & \textbf{12.18 ± 4.93\textsuperscript{*}}
    \end{tblr}
    \medskip
    \caption{Quantitative comparisons of the segmentation quality with and without registration. Mean $\pm$ 95\% Confidence Interval is presented. Better results are shown in \textbf{Bold}. \textsuperscript{*} indicates highly statistically significant evidence ($p<0.001$) to support the difference with the baseline from the two-sided paired Student's \textit{t}-test. \textsuperscript{\textdagger} indicates highly statistically significant evidence to support the difference with the baseline from the Wilcoxon signed-rank test.}
    \label{table: Quantitative Comparisons}
\end{table}
\includeFigure{experiment_1_example}{\useColorTwoPage{Examples of the segmentation of the liver (left half) and pancreas (right half) with and without registration. The pancreas often shows uneven shape and size and its full head-body-tail anatomy is rarely identifiable in a single slice.}}{Experiment 1 Example}

\paragraph{Misalignment Study} We estimated the misalignment within all cases by inverting the annotated transformation parameters, as shown in Figure \ref{figure: Experiment 1 Box Plot}. It can be seen that among all translational offsets, those in the Z-axis show the biggest interquartile range and the most negatively biased distribution. This supports the observation in our prior work \cite{Bagur2021} that translations in the Z-axis are the most prominent source of respiratory motion artifacts during data acquisition irrespective of the organ types. Conversely, rotational offsets are much smaller in magnitude and more randomly distributed compared with the translational ones due to: (1) intuitively, it is hard for patients to rotate their bodies when lying down flat on the scanner; (2) unlike translations, tiny rotations can introduce notable positional changes in the physical space; and (3) in our protocol, rotational motion artifacts are mainly corrected in the fine-tuning phase, where minuscule step sizes are frequently chosen for precise refinement.
\includeFigure{experiment_1_box_plot}{\useColorTwoPage{Graphical comparisons of the estimated translational (left) and rotational (right) misalignment among the dataset. Offsets are defined with respect to the Patient Coordinate System.}}{Experiment 1 Box Plot}

\subsection{Potential Applicability to Clinical Tasks}
\label{subsection: Potential Applicability to Clinical Tasks}
To demonstrate the potential applicability of SVRDA to clinical tasks, we directly integrated it into the MRI quantification of T1, a reliable biomarker used in the early detection and diagnosis of various chronic diseases \cite{Pan2018, Tirkes2016, Messroghli2007}. More specifically, we evaluated whether more consistent quantification on repeated data can be achieved after manual registration compared with the baseline in which no registration was performed.

\paragraph{Dataset} We used 32 pairs of repeated in-house subsets from healthy volunteers, where each subject was scanned by CoverScan \cite{CoverScan} in a test-retest manner. By generating the 3D segmentation labels of both the liver and pancreas for each subset, a dataset of 64 pairs of repeated cases was obtained, each of which was pre-processed following the procedure described in Section \ref{subsection: Indirect Validation via Segmentation}. More detailed parameters for data acquisition can be found in Appendix \ref{subsection: Data Acquisition}.

\paragraph{No Registration \textit{vs.} After Registration} Figure \ref{figure: Experiment 2 Without Noise BA Plot} illustrates the results, demonstrating that SVRDA can be applied to and benefit this clinical task. Although manual registration had no appreciable effects on quantifying liver cases, where no significant misalignment was observed, it considerably improved the consistency in the T1 quantification of the pancreas, resulting in a reduction of both the mean and bias within the T1 discrepancy. Furthermore, the difference in susceptibility of the liver and pancreas to motion artifacts can be realized by comparing the consistency in quantifying these organs without pre-registration. We believe this gap can be predominantly accounted for by the structural and anatomical differences between the liver and pancreas given that all multi-organ data were acquired under the same condition, which coincides with our previous finding in Section \ref{subsection: Indirect Validation via Segmentation}.
\includeFigure{experiment_2_without_noise_ba_plot}{\useColorOnePage{Graphical comparisons of the T1 quantification with and without registration. Top Left: Bland-Altman (BA) plot showing T1 quantification of the liver cases with no registration. Top Right: BA plot showing T1 quantification of the liver cases after registration. Bottom Left: BA plot showing T1 quantification of the pancreas cases with no registration. Bottom Right: BA plot showing T1 quantification of the pancreas cases after registration. Mean $\pm$ 1.96 Standard Deviation (STD) is presented. T1 is quantified by taking the median of pixel values within the output 2D segmentation label.}}{Experiment 2 Without Noise BA Plot}

\paragraph{Challenging Utility} To further challenge the utility of SVRDA, we repeated the experiment after introducing severe motion artifacts to the liver cases, each of which was modeled by a rigid transformation whose parameters were randomly sampled from Gaussian distributions with zero mean and the corresponding standard deviations observed in the previous misalignment study (Section \ref{subsection: Indirect Validation via Segmentation}). As shown in Figure \ref{figure: Experiment 2 With Noise BA Plot},  the quantification quality was notably disrupted by such noise but can be almost recovered to the situation when no misalignment was manually introduced, demonstrating the robustness and reliability of SVRDA under challenging conditions.
\includeFigure{experiment_2_with_noise_ba_plot}{\useColorOnePage{Graphical comparisons of the T1 quantification of the liver cases with and without registration. Top Left: Bland-Altman (BA) plot showing T1 quantification of the vanilla liver cases with no registration. Top Right: BA plot showing T1 quantification of the vanilla liver cases after registration. Bottom Left: BA plot showing T1 quantification of the noisy liver cases with no registration. Bottom Right: BA plot showing T1 quantification of the noisy liver cases after registration. Mean $\pm$ 1.96 Standard Deviation (STD) is presented. T1 is quantified by taking the median of pixel values within the output 2D segmentation label.}}{Experiment 2 With Noise BA Plot}
\section{Conclusion}
\label{section: Conclusion}
We have presented SVRDA, a web-based tool enabling platform-agnostic collaborative annotation of large-scale benchmark datasets for slice-to-volume registration. We strove for user-friendliness and efficiency throughout the system design while ensuring considerable flexibility and extensibility to facilitate future research. We have indirectly validated the effectiveness of SVRDA by evaluating the post-registration segmentation quality and integrated it into a test-retest T1 quantification pipeline to demonstrate its clinical value. Results suggest that SVRDA can serve as an interface for precise motion artifact correction between volumetric and slice data while being potentially applicable to other tasks incorporating 2D/3D image registration.

Despite the encouraging results reported, there remain limitations: First, SVRDA currently only supports rigid transformations with six degrees of freedom. Although this has sufficed to date, more advanced transformations such as affine and non-rigid transformations can be incorporated to cope with more serious misalignment in practice. Second, to ensure low-latency responses, 3D data is resampled onto 2D planes without taking slice thickness into account. Applying a weighted function such as the Gaussian function across the plane thickness would likely improve the accuracy of resampling and interpolation. Finally, SVRDA necessitates complete manual manipulation, which can lead to a laborious user experience when dealing with excessive data. In the future, robust and easy-to-use automatic algorithms should be integrated to facilitate the labeling process by efficiently providing a reliable starting point for further manual refinement.

\section*{Acknowledgements}
\label{section: Acknowledgements}
This research has been conducted using the UK Biobank Resource under application 9914.
\bibliography{reference_list}
\bibliographystyle{unsrtnat}
\section*{Appendix}  
\label{section: Appendix}
\renewcommand{\thesubsection}{\Alph{subsection}}  

\subsection{Registration Protocol}
\label{subsection: Registration Protocol}
Based on \cite{Bagur2021}, our manual registration protocol comprises the following principles and steps:

\paragraph{P1: Follow minimalism.} In line with Occam's razor, we prioritized transformations with notable positive impacts, aiming for optimal outcomes through fewer actions. This effectively minimizes the risk of manually introducing biases by omitting insignificant operations.

\paragraph{P2: Prioritize quality.} We prioritized output quality over annotation efficiency by exclusively operating in Micro Mode to meticulously process each slice.

\paragraph{S1: Correct within-plane misalignment.} We registered $\mathbf{S}_{Selected}$ and $\mathbf{S}_{Resampled}$ mostly through translations in the X and Y-axes, often resulting in more favorable starting points for \textbf{S2}. In this step, we primarily employed Boundary Matching to visually evaluate the alignment.

\paragraph{S2: Correct through-plane misalignment.} We conducted an exhaustive search to determine the optimal alignment by translating $\mathbf{S}_{Selected}$ along the Z-axis until it exhibited the highest similarity with $\mathbf{S}_{Resampled}$ or the alignment was satisfactory.

\paragraph{S3: Fine-tune alignment.} We used small-step transformations to fine-tune the alignment. This was done by applying translations in the X and Y-axes to correct any mismatches in boundaries and features between $\mathbf{S}_{Selected}$ and $\mathbf{S}_{Resampled}$ introduced by previous operations while utilizing rotations to refine the shape of $\mathbf{L}_{2D}$.

\subsection{Data Acquisition}
\label{subsection: Data Acquisition}
Table \ref{table: Acquisition Parameter} summarizes typical acquisition parameters for MRI scans used in the case studies. 
\begin{table}[!ht]
    \centering
    \begin{tblr}{
        column{even} = {c},
        column{3} = {c},
        column{5} = {c},
        column{7} = {c},
        cell{1}{2} = {c=2}{},
        cell{1}{4} = {c=4}{},
        cell{2}{4} = {c=2}{},
        cell{2}{6} = {c=2}{},
        hline{1,4,10} = {-}{},
    }
                             & UK Biobank                &                           & In-house                  &                           &                           &                           \\
                             & 3D VIBE                   & 2D T1                     & 3D VIBE                   &                           & 2D T1                     &                           \\
                             & 1.5T                      & 1.5T                      & 1.5T                      & 3T                        & 1.5T                      & 3T                        \\
    Matrix                   & 320$\times$260$\times$52  & 384$\times$288$\times$7   & 320$\times$260$\times$80  & 224$\times$224$\times$72  & 384$\times$288$\times$1   & 384$\times$288$\times$1   \\
    Resolution (mm$^{2}$)    & 1.19$\times$1.19          & 1.15$\times$1.15          & 1.19$\times$1.19          & 1.96$\times$1.96          & 1.15$\times$1.15          & 1.15$\times$1.15          \\
    Thickness (mm)           & 1.6                       & 8                         & 4                         & 4                         & 8                         & 8                         \\
    TE/TR (ms)               & 1.15/3.11                 & 1.93/480.60               & 2.39/2.38                 & 1.23/1.23                 & 1.97/-                    & 1.97/-                    \\
    Flip Angle (\textdegree) & 10                        & 35                        & 10                        & 9                         & 35                        & 35                        \\
    R                        & 2                         & 2                         & 2                         & 2                         & 2                         & 2                 
    \end{tblr}
    \medskip
    \caption{Summary of acquisition parameters for MRI scans used in the case studies. VIBE: Volumetric Interpolated Breath-hold Examination scan. TE: Time to Echo. TR: Repetition Time. R: Parallel Imaging Acceleration Factor. Scans were acquired from Siemens Aera 1.5/3 Tesla (Siemens Healthineers AG, Erlangen, Germany).}
    \label{table: Acquisition Parameter}
\end{table}

\end{document}